\documentclass[aps,prf,onecolumn,groupedaddress]{revtex4-2}

\usepackage{amsmath}
\usepackage{graphicx}
\usepackage{color}
\usepackage{bm}
\usepackage{graphicx}
\usepackage{epstopdf, epsfig}
\usepackage{amssymb}
\usepackage{tikz}
\usepackage{pgfplots}
\usepackage{soul,xcolor}
\usepackage{natbib}
\usepackage{natbib}
\usepackage{gensymb}
\usepackage{amsmath, bm}
\setstcolor{red}

\newcommand{\tupdo}{\tau_{\uparrow\downarrow}}
\newcommand{\tup}{\tau_{\uparrow}}
\newcommand{\dup}{D_{\uparrow}}
\newcommand{\nup}{n_{\uparrow}}
\newcommand{\jup}{j_{\uparrow}}
\newcommand{\mup}{\mu_{\uparrow}}
\newcommand{\siup}{\sigma_{\uparrow}}
\newcommand{\tdo}{\tau_{\downarrow}}
\newcommand{\ddo}{D_{\downarrow}}
\newcommand{\ndo}{n_{\downarrow}}
\newcommand{\jdo}{j_{\downarrow}}
\newcommand{\mdo}{\mu_{\downarrow}}
\newcommand{\sido}{\sigma_{\downarrow}}

\pdfoutput=1

\begin{document}
	
\title{Electrical voltage by electron spin-vorticity coupling in laminar ducts} 

\author{Hamid Tabaei Kazerooni$^1$, Georgy Zinchenko$^{1,2}$, J\"org Schumacher$^{1,3}$, and Christian Cierpka$^1$}
\affiliation{$^1$ Institute of Thermodynamics and Fluid Mechanics, Technische Universit\"at Ilmenau, 98684 Ilmenau, Germany\\
                $^2$ Moscow Power Engineering Institute, Krasnokazamennaya 14, 111250 Moscow, Russia\\
                $^3$ Tandon School of Engineering, New York University, New York, NY 11201, USA}
\date{\today}

\begin{abstract}
We report a linear scaling law for an electrical voltage as a function of the pressure drop in capillary pipes and ducts. This voltage is generated by a process which is termed spin hydrodynamic generation (SHDG), a result of the collective electron spin--coupling to the vorticity field in the laminar flow in combination with an inverse spin-Hall effect. We study this phenomenon in laminar duct flows with different width-to-height aspect ratios ranging from 1 (square ducts) to infinite (two dimensional channels). First, we analytically solve the governing Valet-Fert spin diffusion equations for the SHDG by means of the method of small parameters together with proper boundary conditions for the set of inhomogeneous elliptic partial differential equations. Secondly, the proposed linear scaling law is validated through a series of experiments using capillary tubes with rectangular and square cross-sections. The experimental results show a very good agreement to the analytically found scaling law. A subsequent substitution of the bulk velocity of the laminar wall-bounded flows by the pressure drop reveals a universal scaling law for the electrical voltage that incorporates all pipe and duct geometries which we could study in our experiments. Finally, the efficiency of the system is estimated for circular pipes, rectangular and square ducts. This study shows that the efficiency of a spin hydrodynamic generator is the same for a circular pipe and a square duct with the same diameter and height, respectively. Hence, due to the ease of manufacturing and the possibility to scale the experiments up to parallel settings in a compact form, micro-channels with a square cross-section seem to be the optimum for a spin hydrodynamic generator. 

\end{abstract}

\maketitle

\section{Introduction}
Spintronics \cite{vzutic2004spintronics} and fluid mechanics, two seemingly unrelated disciplines of physics, have been shown recently to be interconnected through the angular momentum exchange between the local rotation in an electrically conducting fluid flow, which is quantified by the vorticity field, and the internal quantum mechanical angular momentum of the electrons, the electron spin. In their seminal work, \citet{takahashi2016spin} showed that the coupling between these two angular momenta results in the generation of a charge current in the streamwise direction in a turbulent flow of mercury through circular pipe capillaries. To describe this phenomenon, they coined the term \textit{spin hydrodynamic generation} (SHDG) in analogy to the \textit{magnetohydrodynamic generation} which is a well-known conventional technique to convert the kinetic energy of electrically conducting fluids into electricity based on the Lorentz force acting on charged particles in the presence of a magnetic field \cite{Davidson}. In case of the spin hydrodynamic counterpart however, no external magnetic field is required to generate electricity. This makes SHDG a very attractive technique for developing new devices to harvest and directly convert the motion of an electrically conducting fluid into electricity or to measure flow rates in opaque liquid metal flows, to mention two potential fields of technological application. 

The generated electricity is an indirect consequence of the spin current -- `` the flow of electron spins " -- which is induced by a non-equilibrium spin state as a result of the vorticity gradient perpendicular to the flow direction. In case of a non-vanishing spin current in the system under consideration, the so-called inverse spin Hall effect (ISHE)\cite{saitoh2006conversion} is responsible for a charge current. Thus an electrical voltage in the sub-microvolt range will be detectable along the streamwise direction. This direction is perpendicular to that of the spin current. At the moment, experimental studies of SHDG are still in its infancy and scaling laws that describe this generated electrical voltage as a function of flow rates are only available for circular pipe flows \cite{takahashi2016spin, kazerooni2020electron, takahashi2020giant}. In ref. \cite{kazerooni2020electron}, we reported an independent confirmation of the experimental results of \citet{takahashi2016spin}  and were able to extend the range of Reynolds numbers of the flow through circular capillaries with circular cross section to higher and lower values, $20 < Re < 21500$. These experiments have been conducted with the liquid metal alloy galium-indium-tin (in short GaInSn) as a working fluid.  Furthermore, an excellent agreement between the results of refs. \cite{kazerooni2020electron} and very recent investigations of \citet{takahashi2020giant} could be found in the laminar flow regime. This regime is also in the focus of our present work where we vary the cross section geometry.

In the present study, we extend these investigations and report an analytical universal scaling law to predict the generated voltage as a function of the bulk velocity $U_b$ (and thus of the pressure drop $\Delta p$) due to SHDG for this class of laminar rectangular duct flows. Therefore the spin diffusion equation is solved, an inhomogeneous elliptical Helmholtz equation. The analytical solution is obtained by a series expansion with respect to a small parameter that relates the (ab initio unknown) spin diffusion length to the duct half height $b=H/2$. We also provide comprehensive experimental evidence for the reliability of the proposed scaling law using rectangular capillary tubes with square and rectangular cross section of different aspect ratio $W/H$ ($W$ is the width of the duct, $H$ its height). The obtained measurement results are in agreement with the analytical predictions for the electrical voltage $V_{\rm ISHE}$ and display the calculated linear dependence, $V_{\rm ISHE}={\rm const}\times U_b$. Finally, we discuss the efficiency of a spin hydrodynamic generator in case of using circular pipes, square and rectangular ducts. 
 
This paper is organized as follows. In the next section, we describe in brief the theoretical foundations of the SHDG and show how the inhomogeneous version of the Valet-Fert equation of spin diffusion can be used to explain the generated electricity in a liquid metal flow. Section III is dedicated to the analytical solution of this equation for laminar rectangular duct flows. As a results, the universal scaling law for the generated voltage in such geometries will be presented. Section IV of the paper is devoted to the verification of the proposed scaling law by means of experiments and also estimating the SHDG efficiency of different capillary tubes using measured voltages and mean flow velocities. We describe the setup and summarize the measurement results. A short discussion of the efficiency of this method is also provided before we conclude and give a brief outlook into future activities on this subject. 

\begin{table*}[t]
\renewcommand{\arraystretch}{1.4}
\centering
\begin{tabular}{ccccc}
\hline\hline
                 &  $\quad$ &   Electronics	    & $\quad$&   Spintronics \\
\hline
Transported quantity  & &  $e^-$      & & $s_z=\pm \dfrac{\hbar}{2}$  \\
Current density          & &  $j_k^C$  & & $j_{kl}^S$ \\
Voltage                      & &  $\Phi^C$ & &  $\mu_k^S$ \\
Ohm's law                 & &  $j_k^C = -\sigma_0 \dfrac{\partial \Phi^C}{\partial x_k}$ & &  $j_{kl}^S = -\dfrac{\hbar \sigma_0}{4 e^2} \dfrac{\partial \mu_l^S}{\partial x_k}$\\
\hline
\hline
\end{tabular}
\caption{Analogy of central physical quantities in electronics and spintronics. Indices $k,l=1,2,3$. Quantity $\sigma_0$ is the electrical conductivity, $\hbar=h/{2\pi}$  is the smallest unit of the angular momentum containing Planck's constant $h$, and $e$ is the elementary charge. Spin-up corresponds to $s_z=+\hbar/2$, spin-down to $s_z=-\hbar/2$. We list the definitions of both current densities by an Ohm's law. Since the electron spin can be considered as a vectorial quantity, the related current density has to be tensorial. }
\label{Tab1}
\end{table*}
 
\section{Collective electron spin diffusion}
\subsection{Spin current and spin voltage}
We start with a brief overview on the basic principles that cause the spin hydrodynamic generation in the liquid metal flow. Table \ref{Tab1} relates central physical quantities such as current (density), voltage, and their connection via Ohm's law of electronics to corresponding ones in spintronics. Similar to the vectorial charge current $j_k^C$ in electronics, the generation and control of the tensorial spin current $j_{kl}^S$  are the main concern in spintronics \cite{vzutic2004spintronics,hirohata2014future}. However, these tasks are difficult as the latter is not conserved in contrast to its electronic counterpart. Nevertheless, different methods have been employed to induce a spin current based on an angular momentum exchange as it will also be the case for the present setting. We mention here the exchange between spins and light polarization \cite{ando2010direct}, a temperature gradient \cite{uchida2008observation}, or mechanical rotation of a material system \cite{matsuo2013mechanical}. Both current densities are coupled by the inverse Hall spin effect (ISHE), that converts a spin current into an electric current. Therefore the electrons are deflected perpendicular to the direction of their flow depending on the sign of their spin which leads to  
\begin{equation}
\label{charg_curr1}
j_{k}^{C}=-\frac{2|e|}{\hbar} \theta_{\mathrm{SHE}}\, \epsilon_{klm}\, j_{lm}^{S}\,,
\end{equation}
where $\epsilon_{klm}$ is the fully anti-symmetric third rank Levi-Civita tensor and $\theta_{\mathrm{SHE}}$ is the spin-Hall angle, a material-dependent constant \cite{saitoh2006conversion}. We apply the Einstein summation convention. The local rotation in the fluid is quantified by the vorticity field $\omega_{k}=\epsilon_{klm} \partial u_m/\partial x_l$. The indices are always taken as  $k,l,m=1,2,3$. The coupling of vorticity and spin voltage is established in a spin diffusion equation  which will be discussed in the next subsection. 

\subsection{Spin diffusion and the Fert-Valet equation}
At the core of the theoretical modeling is a diffusion dynamics of the spin voltage which contains an important length scale, the spin diffusion length $\lambda$. In analogy to rarified gases, one can formulate a kinetic gas theory for the electron transport in solid and liquid conductors. Here, we discuss the one-dimensional case for simplicity that leads to a diffusion dynamics. The Ohm's law which we listed in table I can be extended by a drift term if the number density $n$ of the electrons varies in space,    
\begin{equation}
\label{charg_curr3}
j^{C}= -\sigma_0 \frac{\partial \Phi}{\partial x}  -e D \frac{\partial n^C}{\partial x}= \frac{\sigma_0}{e} \frac{\partial}{\partial x}\left[-e\Phi - \frac{e^2 D}{\sigma_0} n^C\right] = \frac{\sigma_0}{e} \frac{\partial \mu}{\partial x}\,.
\end{equation}
Here, $D$ is a diffusion constant and $\mu$ is denoted to as the electrochemical potential. Now we decompose the electron ensemble into spin-up and spin-down fractions with $\nup$, $\jup$, $\mup$, $\siup$, $\dup$ and $\ndo$, $\jdo$, $\mdo$, $\sido$, $\ddo$ and define the following number and currents densities, leaving aside the vectorial or tensorial character for the moment,
\begin{equation}
\label{charg_curr4}
n^{C}= \nup + \ndo\,,\quad n^S=\nup-\ndo\,, \quad \text{and} \quad j^{C}= \jup + \jdo\,,\quad j^S=\jup-\jdo\,.
\end{equation}
Charge conservation is in line with a divergence-free electrical current density and thus with \eqref{charg_curr3} one gets
\begin{equation}
\label{charg_curr5}
0= \frac{\partial j^C}{\partial x} = \frac{\siup}{e}\frac{\partial^2\mup}{\partial x^2} +  \frac{\sido}{e}\frac{\partial^2\mdo}{\partial x^2}\,.
\end{equation}
Spin flips from up to down or vice versa proceed on a time scale which is larger than a typical relaxation time, $\tupdo\gg\tup, \tdo$. The spin conservation is captured by the following relation, for which we use $j=e n v_{\rm drift}$ in general, 
\begin{equation}
\label{charg_curr6}
\frac{\partial \jup}{\partial x} - \frac{\partial \jdo}{\partial x} = e \left(\frac{\nup - \ndo}{\tupdo} \right)\,.
\end{equation}
A combination of eqns. \eqref{charg_curr4}, \eqref{charg_curr5} and \eqref{charg_curr6} leads to a relation for the electrochemical potentials of both fractions,
\begin{equation}
\label{charg_curr7}
\frac{\partial^2 \mup}{\partial x^2} = \frac{e^2}{2\siup} \left(\frac{\nup - \ndo}{\tupdo} \right)\,, \quad \frac{\partial^2 \mdo}{\partial x^2} = -\frac{e^2}{2\sido} \left(\frac{\nup - \ndo}{\tupdo} \right)\,,
\end{equation}
and thus to 
\begin{equation}
\label{charg_curr7a}
\frac{\partial^2 \mup}{\partial x^2}-\frac{\partial^2 \mdo}{\partial x^2}=\left(\frac{e^2}{2\siup}+ \frac{e^2}{2\sido}\right)\, \frac{2\nup}{\tupdo}\,.
\end{equation}
We have used the fact, that a screening of charges is effective over microscopic lengths such that $\nup\approx \ndo$. Together with the definition of electrochemical potential (first term in \eqref{charg_curr3} cancels) one gets
\begin{equation}
\label{charg_curr8}
\mup-\mdo = \left(\frac{e^2\dup}{\siup}+ \frac{e^2\ddo}{\sido}\right)\,\nup\,.
\end{equation}
We can finally combine eqns. \eqref{charg_curr7a} and \eqref{charg_curr8} to a relation for the potential difference
\begin{equation}
\label{charg_curr9}
D_F\left(\frac{\partial^2 \mup}{\partial x^2}-\frac{\partial^2 \mdo}{\partial x^2}\right)=\left(\frac{\mup-\mdo}{\tupdo}\right) \quad\text{with}\quad D_F=\frac{\siup\ddo+\sido\dup}{\siup+\sido}
\,.\end{equation}
Here $D_F$ is an effective diffusion constant composed of both sub-ensembles. In analogy to the classical diffusion, we can define a {\em spin diffusion length} -- the characteristic length scale between a spin-flip by $\lambda=\sqrt{D_F\,\tupdo}$ (similar to a mean free path between two gas particle collisions). The potential difference $\mu^S=\mup-\mdo$ is denoted as spin voltage and \eqref{charg_curr9} gets its final form suggested first by Valet and Fert \cite{valet1993theory}
\begin{equation}
\label{charg_curr10}
\frac{\partial^2 \mu^S}{\partial x^2}= \frac{1}{\lambda^2}\mu^S\,.
\end{equation}
This is a linear elliptic equation of second order. The coupling to the local vorticity field enters this equation as an additional inhomogeneity on the right hand side and for 3 dimensions follows \cite{saitoh2006conversion,takahashi2016spin}
\begin{equation}
\label{main_spv}
\nabla^{2} \mu_k^{S}=\frac{1}{\lambda^{2}} \mu_k^{S}-\frac{4 e^{2} \xi}{\sigma_{0} \hbar} \omega_k\,.
\end{equation}
Here, $\xi$ is defined by \citet{takahashi2016spin} as a parameter which represents the angular momentum transfer from the fluid to the spins. Note also that $\mu_k^{S}$ is aligned with the flow vorticity --  the quantity that determines the magnitude of the spin voltage. Quantities $\lambda$, $e$, $\hbar$ and $\sigma_{0}$ are the spin diffusion length, the elementary charge, the reduced Planck constant and the electrical conductivity of the working liquid metal, respectively. Equation \eqref{main_spv} is solved analytically in the following section for the laminar duct case to determine the measurable electrical voltage $V_{\rm ISHE}$.

\section{Analytical solution of the spin diffusion equation for duct flow}
\begin{figure}[h!]
	\centering
	\includegraphics[width=0.7\textwidth,trim={0cm 1cm 0cm 1cm},clip]{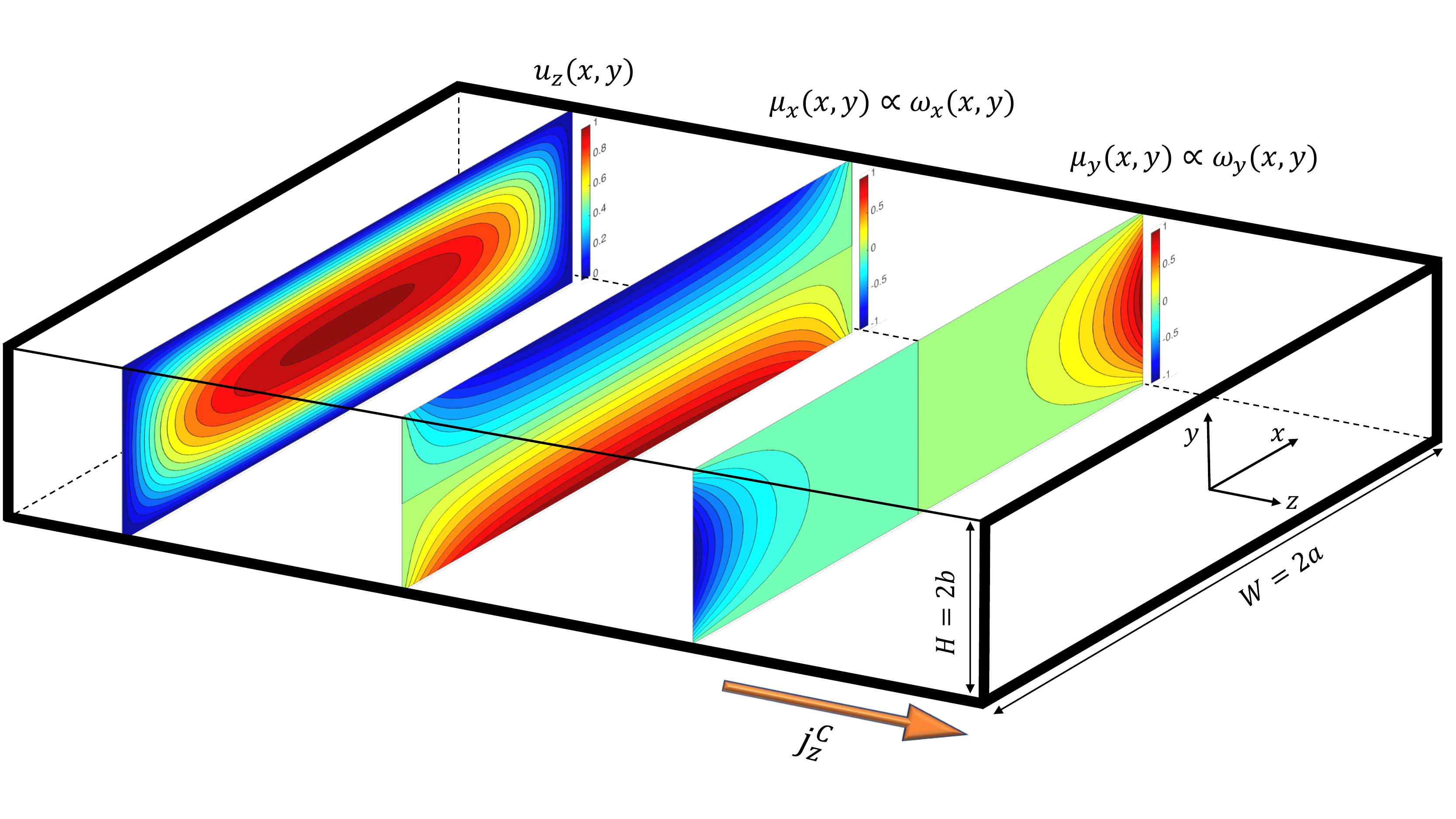}
	\caption{Schematic of the rectangular duct geometry with aspect ratio $W/H=3$. Contour plots of the streamwise velocity profile $u_z(x,y)$ and the in-plane spin voltages $\mu_x(x,y)$ and $\mu_y(x,y)$ are also indicated in the figure. Due to the inverse spin hall Effect (ISHE), the in-plane spin current $j_{xy}^{\mathrm{S}}$ induces a charge current $j_{z}^{\mathrm{C}}$ in the flow direction. Data are normalized with the maximum value of each quantity.}
	\label{Rec}
\end{figure}

We now obtain the scaling law for the generated voltage $V_{\rm ISHE}$ from the Valet-Fert equation \cite{valet1993theory} for a laminar pressure-driven rectangular duct flow with different aspect ratios $W/H$ where the duct width and height are presented as $W=2a$ and $H=2b$ in a Cartesian coordinate system with its origin located at the center of the cross section and $-a\le x\le a$ and $-b\le y\le b$ (see Fig. \ref{Rec}).  One has to solve the Stokes flow problem for no-slip boundary conditions at the walls. The vorticity components are taken from the streamwise velocity profile $u_{z}(x,y)$ in a rectangular duct which is given by (see e.g. ref. \cite{shah2014laminar}),  
\begin{equation}
\label{velocity_rectangle}
u_{z}(x,y) = \frac{1}{\mu} \frac{\Delta p}{L} \sum_{n=1,3,...}^{\infty} \frac{(-1)^{\frac{n-1}{2}}}{(\pi n)^{3}} 16b^2 \left[1-\frac{\cosh \left(\frac{\pi n x}{2b}\right)}{\cosh \left(\frac{\pi n a}{2 b}\right)}\right] \cos \left(\frac{\pi ny}{2b}\right), 
\end{equation}
and follow to
\begin{align}
\label{omega_x}
\omega_{x}(x, y)&=\frac{\partial u_{z}}{\partial y} = -\frac{1}{\mu} \frac{\Delta p}{L} \sum_{n=1,3,...}^{\infty} \frac{(-1)^{\frac{n-1}{2}}}{(\pi n)^{2}} 8b \left[1-\frac{\cosh \left(\frac{\pi n x}{2b}\right)}{\cosh \left(\frac{\pi n a}{2 b}\right)}\right] \sin \left(\frac{\pi n y}{2b}\right)\,,\\ 
\label{omega_y}
\omega_{y}(x, y)&=-\frac{\partial u_{z}}{\partial x} = \frac{1}{\mu} \frac{\Delta p}{L} \sum_{n=1,3,...}^{\infty} \frac{(-1)^{\frac{n-1}{2}}}{(\pi n)^{2}} 8b \frac{\sinh \left(\frac{\pi n x}{2b}\right)}{\cosh \left(\frac{\pi n a}{2 b}\right)} \cos \left(\frac{\pi n y}{2b}\right)\,.
\end{align}
Using $u_{z}(x,y)$ and the flow rate $Q$, the bulk velocity $U_b$ is given by,
\begin{align}
U_b&=\frac{Q}{A}=\frac{1}{4ab}\int_{-b}^{b} \int_{-a}^{a} u_{z}(x,y) dxdy\nonumber \\
&=-32 b^{2} \frac{1}{\mu} \frac{\Delta p}{L} \sum_{n=1,3,...}^{\infty}\left[\frac{1}{(\pi n)^{4}}-\frac{2}{(\pi n)^{5}} \frac{b}{a} \tanh \left(\frac{\pi n a}{2 b}\right)\right]\,.
\end{align}
For $a \geq b$ , we use $\tanh (\pi na/2b)\approx 1$ and the series terms can be estimated using the Riemann Zeta function $\zeta(s)$ \cite{karatsuba2011riemann}. Hence, the pressure drop $\Delta p$ and the flow bulk velocity $U_b$ in a rectangular duct flow are related via the following approximation,
\begin{equation}
\frac{1}{\mu} \frac{\Delta p}{L}\approx\frac{3 U_{b}}{b^{2}} \frac{1}{1-\frac{186}{\pi^{5}} \zeta(5) \frac{b}{a}} \quad\mbox{with}\quad a \geq b\,.
\end{equation}
Using the above expression, eqns.~(\ref{velocity_rectangle})--(\ref{omega_y}) can be rewritten on the basis of the flow bulk velocity $U_b$ which can be easily measured experimentally. With the known vorticity profiles for the rectangular duct flow, we can now solve the inhomogeneous spin diffusion equation \eqref{main_spv}. For simplicity, we switch to the dimensionless form of eq.~(\ref{main_spv}) using the following parameters: the half-height $b$ for lengths, the characteristic spin voltage $\mu_{ch}$, and the bulk velocity $U_b$ for velocity. Utilizing these parameters and defining, 
\begin{equation}
\Gamma^{2}=\frac{b^{2}}{\lambda^{2}} \quad\mbox{and}\quad \mu_{ch}=U_b\frac{4 e^{2}}{\sigma_{0} \hbar} \xi b\,,
\end{equation}
eq.~\eqref{main_spv} is converted to the following inhomogeneous elliptic equation
\begin{equation}
\label{main_spv_ND_2}
\nabla^2 \mu_k^{\mathrm{S}}=\Gamma^{2} \mu_k^{\mathrm{S}}-\epsilon_{klm} \frac{\partial u_m}{\partial x_l}\,.
\end{equation}
To solve eq.~(\ref{main_spv_ND_2}) for a laminar rectangular duct flow, we employ the method of small parameter for differential equations \cite{wasow2018asymptotic}. We check the reliability of the method first by considering the simple case of SHDG in a steady laminar flow between two infinitely long parallel plates with a possible analytical solution. The well-known velocity profile for a laminar Poiseuille channel flow reads in dimensionless form as $u_z(y)=(3/2)(1-y^2)$. Thus $\omega_{x}=-3y$ and $\omega_{y}=0$ follow. Given the physical fact that there is no spin current through the channel walls at top and bottom, i.e.,  $j^{\mathrm{S}}_{kl} \propto \partial \mu_k^{\mathrm{S}}/\partial x_l=0$, and no spin voltage at the channel center due to the symmetry, $\mu_k^{\mathrm{S}}(0)=0$, eq.~(\ref{main_spv_ND_2}) simplifies in the channel flow case to
\begin{equation}
\label{ODE_Channel}
\frac{\partial^{2} \mu_{x}^{\mathrm{S}}}{\partial y^{2}}=\Gamma^{2} \mu_{x}^{\mathrm{S}}+3 y \quad\mbox{with}\quad \frac{\partial \mu_{x}^{\mathrm{S}}}{\partial y}\Big|_{y=\pm 1}=0,\;  \mu_{x}^{\mathrm{S}}(y=0)=0\,.
\end{equation}
The solution of this equation consists of a homogeneous and an inhomogeneous part, i.e. $\mu_{x}^{\mathrm{S}}=\mu_{x,\mathrm{h}}^{\mathrm{S}}+\mu_{x,\mathrm{inh}}^{\mathrm{S}}$ with $\mu_{x,\mathrm{inh}}^{\mathrm{S}}=-3y/\Gamma^{2}$ and $\mu_{x,\mathrm{h}}^{\mathrm{S}}=A \sinh (\Gamma y)+B \cosh (\Gamma y)$ where $B=0$ due to the second boundary condition at the center, i.e. $\mu_{x}^{\mathrm{S}}(0)=0$. From the first boundary condition at the walls one obtains, $A={3}
/{(\Gamma^3 \cosh(\Gamma))}$. Thus, the spin voltage distribution in a laminar channel flow is given by
\begin{equation}
\label{Analy_Chan}
\mu_{x}^{\mathrm{S}}(y)=\frac{3}{\Gamma^{3}}\left[\frac{\sinh (\Gamma y)}{\cosh (\Gamma)}-\Gamma y\right]
\end{equation}
Now we consider the case of $\Gamma \gg 1$ and discuss an approximate solution which is obtained by an expansion ansatz for (\ref{main_spv_ND_2}) which is compared then to the analytical solution (\ref{Analy_Chan}). It should be noted that $\Gamma \gg 1$ is indeed a valid assumption as in practice the spin diffusion length $\lambda \sim 10^{-8}$ m \cite{bass2007spin} is at least two order of magnitude smaller than the capillary tube characteristic length $b$. The ansatz for the solution reads as the following series expansion \cite{wasow2018asymptotic}:
\begin{equation}
\label{ODE_Channel_decom}
\mu_{x,app}^{\mathrm{S}}(y)=\sum_{i=0}^{\infty} \frac{\mu_{i}}{\Gamma^{2i}}\,.
\end{equation}
This series expansion is plugged into the inhomogeneous ordinary differential equation \eqref{ODE_Channel} and will be truncated after the first $N$ terms (see below). Thus follows
\begin{equation}
\frac{\mu_{0}^{\prime \prime}}{\Gamma^{2}}+\frac{\mu_{1}^{\prime \prime}}{\Gamma^{4}}+\ldots+\frac{\mu_{N}^{\prime \prime}}{\Gamma^{2 N+2}}=\mu_{0}+\frac{\mu_{1}}{\Gamma^{2}}+\ldots+\frac{\mu_{N}}{\Gamma^{2 N}}+\frac{3 y}{\Gamma^{2}}\,.
\end{equation} 
Sorting with respect to powers of $\Gamma$ and stopping at $N=2$, one obtains the following equations
\begin{equation}
\mu_{0} =0\,,  \quad\quad
\frac{\mu_{0}^{\prime \prime}}{\Gamma^{2}} =\frac{\mu_{1}+3 y}{\Gamma^{2}}\,, \quad\quad
\frac{\mu_{1}^{\prime \prime}}{\Gamma^{4}} =\frac{\mu_{2}}{\Gamma^{4}}\,.
\end{equation}
Hence, the approximated solution follows to
\begin{equation}
\label{app_Chanel_solution}
\mu_{1}=-3 y, \quad \mu_{0}=\mu_{2}=\mu_{3}=\ldots=0 \quad\Rightarrow\quad \mu_{x,app}^{\mathrm{S}}(y)=-\frac{3y}{\Gamma^{2}}\,,
\end{equation}
which implies that for $\Gamma \gg 1$, the first term in eq.~(\ref{Analy_Chan}) converges to zero and both eqns.~(\ref{Analy_Chan}) and (\ref{app_Chanel_solution}) agree. Next, we apply the same procedure to solve eq.~(\ref{main_spv_ND_2}) for a laminar rectangular duct flow with considering the proper boundary conditions:
\begin{equation}
\label{app_duct_eq}
\frac{\partial^2 \mu_k^{\mathrm{S}}}{\partial x^2}+\frac{\partial^2 \mu_k^{\mathrm{S}}}{\partial y^2} =\Gamma^{2} \mu_k^{\mathrm{S}}-\omega_k \quad\mbox{with}\quad
\frac{\partial \mu_k^{\mathrm{S}}}{\partial x}\Bigg|_{x=\pm \frac{a}{b}}=0, \; 
\frac{\partial \mu_k^{\mathrm{S}}}{\partial y}\Bigg|_{y=\pm 1}=0, \; 
\mu_k^{\mathrm{S}}(x=0,y=0)=0\,.
\end{equation}
It should be mentioned again that here the spin voltage is a vectorial quantity (and thus $k=x,y$) and that (\ref{app_duct_eq}) has to be solved for each component of the vorticity $\omega_k$ as given by eqns. (\ref{omega_x}) and (\ref{omega_y}). Now the series expansion \eqref{ODE_Channel_decom} is the only possible way to advance to an analytical solution which we discuss in the following. The series expansion \eqref{ODE_Channel_decom} is done for each component of the spin voltage and inserted in \eqref{app_duct_eq}. Thus the voltages $\mu_{k,i}$ appear in the series expansion. The following hierarchy of equations follows for $i\le N=3$
\begin{equation}
\label{seri_duct_spin}
\left\{\begin{array}{l}
\displaystyle \mu_{k,0}=0 \\[10pt]
\displaystyle \frac{\mu_{k,1}}{\Gamma^{2}}=\frac{\omega_k(x,y)}{\Gamma^{2}} \\[10pt]
\displaystyle \frac{\nabla^2 \mu_{k,1}}{\Gamma^{4}}=\frac{\mu_{k,2}}{\Gamma^{4}} \\[10pt]
\displaystyle \frac{\nabla^2 \mu_{k,2}}{\Gamma^{6}}=\frac{\mu_{k,3}}{\Gamma^{6}}
\end{array} \Rightarrow\left\{\begin{array}{l}
\displaystyle \Gamma^{-2} \mu_{k,1}=\Gamma^{-2} \omega_k(x,y) \\
\displaystyle \Gamma^{-4} \mu_{2k,}=\Gamma^{-4} \nabla^2 \omega_k(x,y) \quad \quad\Rightarrow \Gamma^{-2i} \displaystyle \mu_{k,i}=\Gamma^{-2 i} (\nabla^2)^{i-1} \omega_k(x,y) \\
\displaystyle \Gamma^{-6} \mu_{k,3}=\Gamma^{-6} \nabla^2 \,\nabla^2 \omega_k(x,y)
\displaystyle \end{array}\right.\right.
\end{equation} 
The vectorial spin voltage in a rectangular duct flow follows to
\begin{equation}
\label{vect_Spin_rec}
\mu_k^{\mathrm{S}}=\sum_{i=1}^3(\nabla^2)^{i-1} \omega_k(x,y)\,.
\end{equation}
Clearly, in order to consider more than one term in the above series solution, the vorticity profiles of eqns.(\ref{omega_x}) and (\ref{omega_y}) need to be differentiated in their non-dimensional form as:~
\begin{equation}
\label{vort_diff}
\begin{aligned}
\nabla^2 \omega_{x} &=\frac{6}{1-\frac{186}{\pi^{5}} \zeta(5)\frac{b}{a}}~~\sum_{n=1,3,...}^{\infty}(-1)^{\frac{n-1}{2}} \sin \left(\frac{\pi n y}{2}\right) \\
\nabla^2 \omega_{y} &=0
\end{aligned}
\end{equation}
We can see that $\mu_{y, 2}=\nabla^2 \omega_{y}=0$ and $\mu_{x,2}=\nabla^2 \omega_{x}=0$ for $y \neq \pm 1$. Hence, there exits only one term ($i=1$) in the approximate solution of eq.~(\ref{vect_Spin_rec}) which can be expanded as,
\begin{equation}
\label{Spin_Volt_rec}
\mu_{x}^{\mathrm{S}}(x,y) =\Gamma^{-2} \mu_{x,1}=\Gamma^{-2} \omega_{x}(x,y) \quad\mbox{and}\quad
\mu_{y}^{\mathrm{S}}(x,y) =\Gamma^{-2} \mu_{y,1}=\Gamma^{-2} \omega_{y}(x,y)\,.
\end{equation}
This shows that the spin voltage in each direction is directly proportional to the corresponding flow vorticity in that direction (see Fig. \ref{Rec}). Now, we estimate the generated in-plane spin $j_{xy}^{\mathrm{S}}$ and the streamwise electric $j_{z}^{\mathrm{C}}$ currents in a laminar rectangular duct flow based on the inverse spin Hall effect \cite{saitoh2006conversion}
\begin{equation}
\label{Spin_Hall}
j_{kl}^{\mathrm{S}}=-\frac{\hbar \sigma_{0}}{4e^{2}} \frac{\partial \mu_l^{\mathrm{S}}}{\partial x_k}
\end{equation}
as we have listed in Table I already. Following eq.~\eqref{charg_curr1} the induced electric current in the streamwise direction, $j_{z}^{\mathrm{C}}$, follows to
\begin{equation}
j_{z}^{\mathrm{C}} (x,y)=\frac{\sigma_{0}}{2\left|e\right|} \theta_{\mathrm{SHE}} \frac{\mu_{c h}}{b}\left(\frac{\partial \mu_{x}^{\mathrm{S}}}{\partial y}-\frac{\partial \mu_{y}^{\mathrm{S}}}{\partial x}\right)
\end{equation}
Differentiating eqns.~(\ref{Spin_Volt_rec}) and substituting them in the above expression, the excited electric current in the system becomes
\begin{equation}
j_{z}^{\mathrm{C}} (x,y)=-\frac{\sigma_{0}}{\left|e\right|} \theta_{\mathrm{SHE}} \frac{\mu_{ch}}{b} \frac{1}{\Gamma^{2}} \frac{6}{1-\frac{186}{\pi^{5}} \zeta(5)\frac{b}{a}}~~\sum_{n=1,3,...}^{\infty} \frac{(-1)^{\frac{n-1}{2}}}{(\pi n)} \cos \left(\frac{\pi n y}{2}\right)
\end{equation}
where the series part converges to a constant value of 1/4. Therefore, the mean charge current $\left\langle j_{z}^{\mathrm{C}}\right\rangle$ follows to
\begin{equation}
j_{z}^{\mathrm{C}} (x,y)=\left\langle j_{z}^{\mathrm{C}}\right\rangle=\frac{b}{4a} \int_{-a / b}^{a / b} \int_{-1}^{1} j_{z}^{\mathrm{C}}(x,y)~dy dx=-\frac{\sigma_{0}}{2\left|e\right|} \theta_{\mathrm{SHE}} \frac{\mu_{ch}}{b} \frac{1}{\Gamma^{2}} \frac{3}{1-\frac{186}{\pi^{5}} \zeta(5)\frac{b}{a}}
\end{equation}
Considering the relation between electrical voltage and resistance,  $V_{\mathrm{ISHE}}=R \, A \, \left\langle j_{z}^{\mathrm{C}}\right\rangle$ with $R=L/(\sigma_{0}A)$, the length $L$ and the cross section $A$ 
\begin{equation}
\label{scaling_rect}
\frac{b^{3}V_{\mathrm{ISHE}}}{L}=\frac{6|e|}{\hbar} \, \frac{\theta_{\mathrm{SHE}} \lambda^{2} \xi_{\mathrm{lam}}^\mathrm{Rec}}{\sigma_{0}} \, \frac{1}{1-\frac{186}{\pi^{5}} \zeta(5) \frac{b}{a}} \, U_b~b
\end{equation}
A similar relation can be derived for  the pipe flow case (with different prefactors due to geometry) which is given in ref. \cite{kazerooni2020electron} for the laminar flow case. We stress at the end of this section that the spin diffusion length $\lambda$, the spin Hall angle $\theta_{\rm SHE}$, and the angular momentum transfer parameter $\xi_{\rm lam}$  cannot be obtained by a macroscopic fluid dynamical consideration. They have to be determined experimentally for the corresponding conductor ($\lambda$, $\theta_{\rm SHE}$) and estimated by linear response theory ($\xi$) assuming white-in-time vorticity fields. Here, the slope of $V_{\rm ISHE}={\rm const}\times U_b$ is obtained directly from the measurements and the product of the three unknowns $\theta_{\mathrm{SHE}} \lambda^{2} \xi_{\mathrm{lam}}^\mathrm{Rec}$ is a free fit parameter in the analytical model.   

\section{Measurements in ducts of different cross sections}
\subsection{Experimental setup and measurement results}
\begin{figure}[h]
	\centering
	\includegraphics[width=0.6\textwidth,trim={0cm 2cm 15cm 0cm},clip]{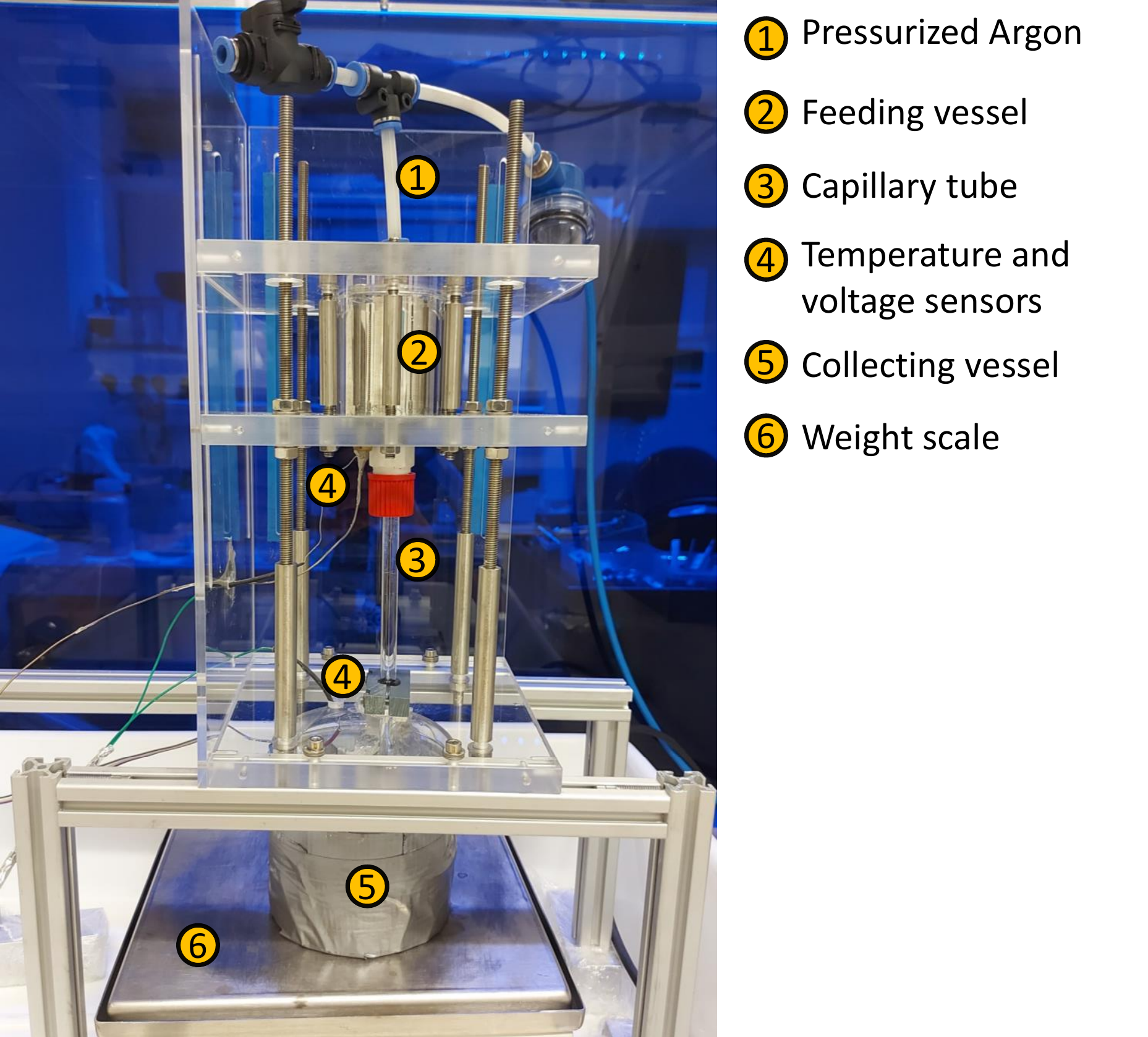}
	\caption{Experimental set-up for measuring the generated electrical voltage $V_{\rm ISHE}$ in a capillary pipe or duct due to the spin hydrodynamic generation. The essential components of the experiment are indicated in the photography and given in the legend.}
	\label{setup}
\end{figure}

We employ the same apparatus and working liquid -- the eutectic alloy GaInSn -- as in our previous experimental study \cite{kazerooni2020electron} where we also discussed in detail that thermoelectric effects remain subdominant and that a dynamo action can be excluded for the voltage generation.  Here, we measure the generated voltage in capillary ducts with different aspect ratios of $W/H=1, 2.34, 10, 20$ at low Reynolds numbers $Re=U_b H/\nu <1400$ where the flow is fully laminar. Figure \ref{setup} shows a photo of the experimental setup. The flow is generated in the capillary tube (3) by pushing the liquid metal in the top vessel (2) using pressurized Argon gas (1). Here, we use glass capillary ducts (VitroCom and Hilgenberg GmbH) with 200 mm length $L$ and sectional dimensions of $H \times W= 0.2\times 0.2 , 0.3\times 0.3, 0.149\times 0.334, 0.1\times 1, 0.05\times 1~\mathrm{mm}^2$. The flow rate is estimated by measuring the weight change of the collecting vessel (5) by means of an accurate weight scale (6). The generated voltage and the working liquid temperature are monitored by the sensors implemented in the top and the bottom vessels (4). Further detailed information about the experimental procedure, data acquisition and the working liquid properties can be found in \cite{kazerooni2020electron}. 

\begin{figure}[h]
	\centering
	\includegraphics[width=0.45\textwidth,trim={0cm 0cm 0cm 0cm},clip]{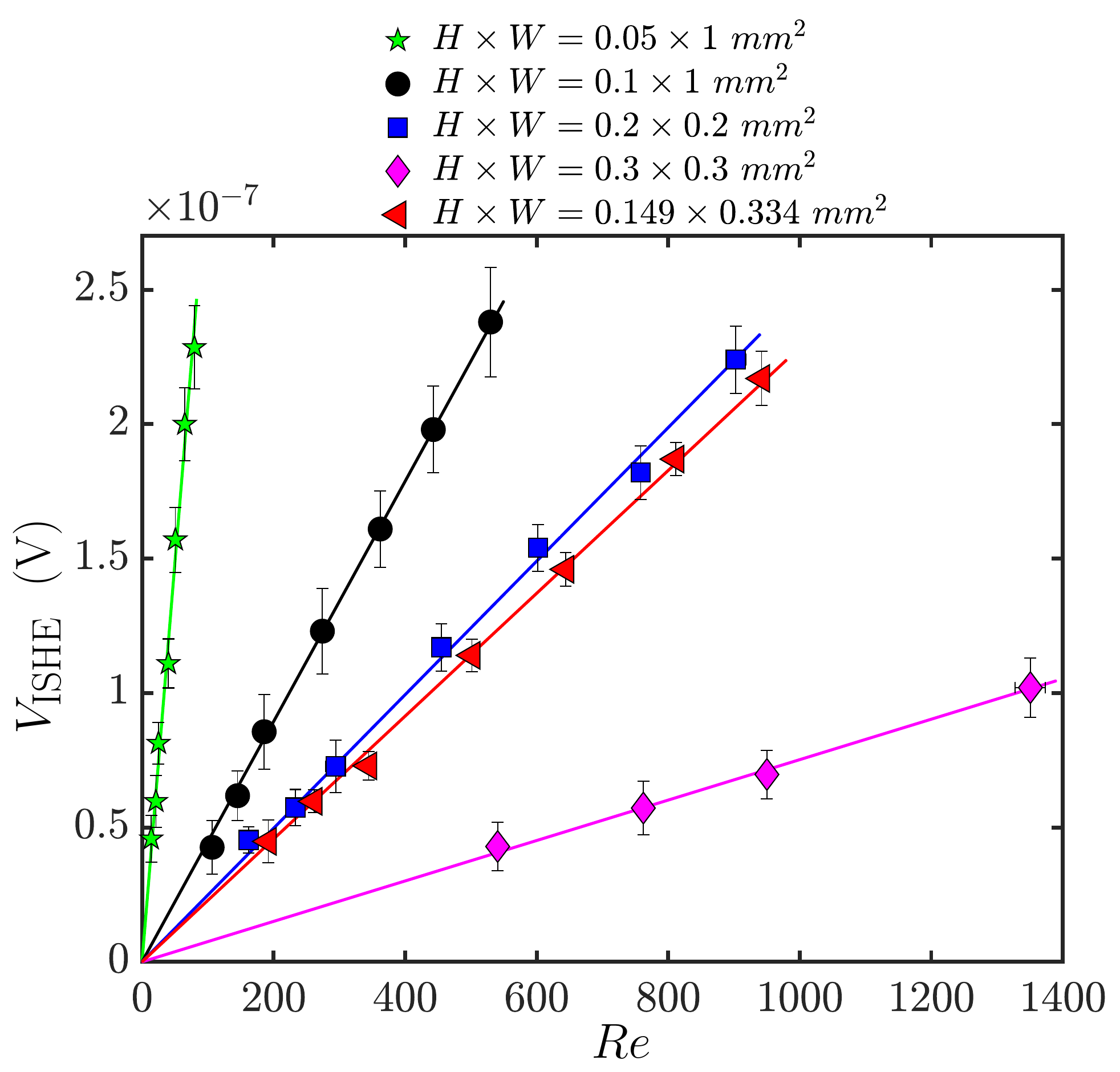}
	\caption{The generated electrical voltage $V_{\mathrm{ISHE}}$ versus the flow Reynolds number $Re$ in square and rectangular capillary tubes. The error bars show the standard deviation of at least five independent measurements.}
	\label{Fig_Volt}
\end{figure}

\begin{figure}[h!]
	\centering
	\includegraphics[width=0.9\textwidth,trim={0cm 0cm 0cm 0cm},clip]{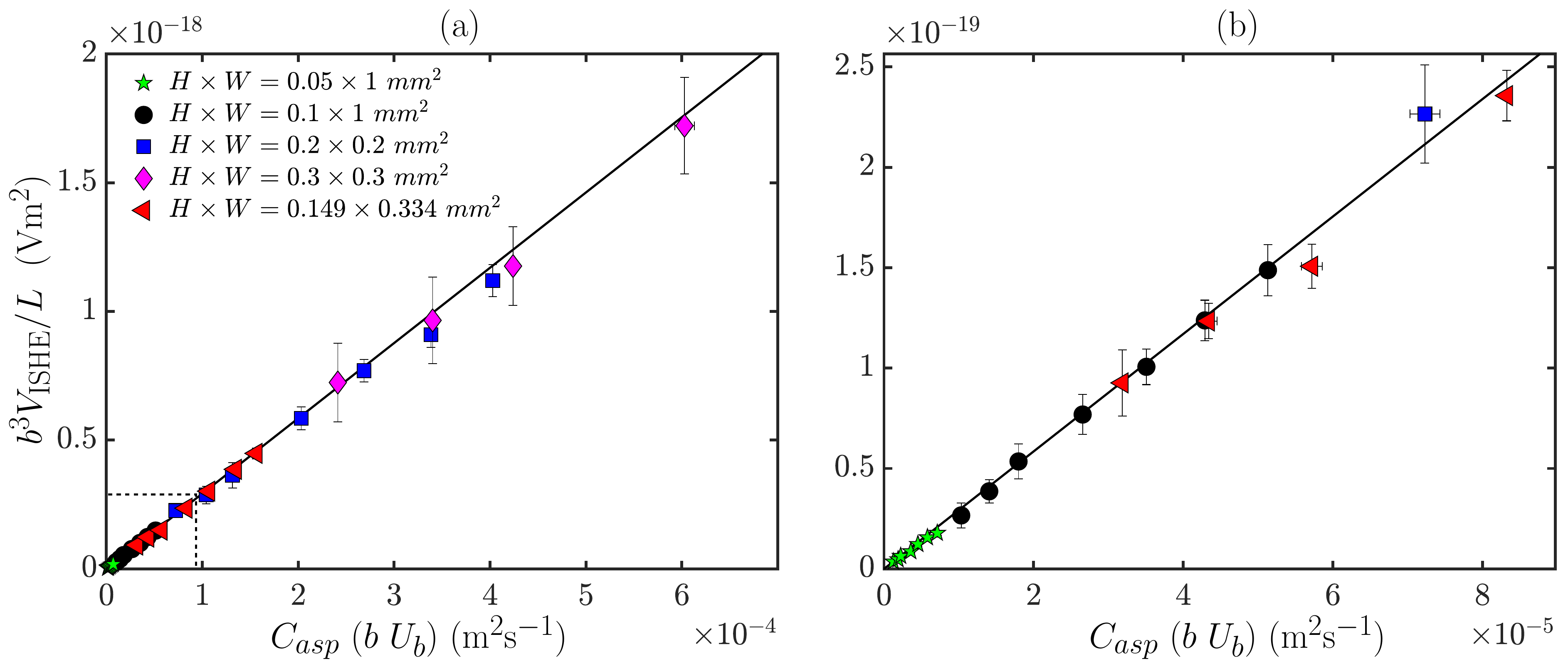}
	\caption{The generated voltage $V_{\mathrm{ISHE}}$ scaled based on eq.~(\ref{scaling_rect}) with respect to $C_{asp}~(bU_b)$ where $C_{asp}=1/(1-(186/\pi^{5})\times(\zeta(5)~b/a))$ is the geometrical factor (a) the full measurement range, (b) data points in the small area indicated by the dashed lines in (a). The solid lines in both panels are fits of the analytical solution to the experimental data.}
	\label{Fig_scale}
\end{figure}

Figure. \ref{Fig_Volt} shows the measured voltages for different square and rectangular cross sections with respect to the flow Reynolds number $Re$. The linear evolution of the voltage can be clearly seen for all cases presented in this figure. The rescaled data points based on eq.~(\ref{scaling_rect}) are presented in Fig.~\ref{Fig_scale}(a) where all collapse into a single linear curve. Fig.~\ref{Fig_scale}(b) shows the data points located in the area indicated by the dashed lines in Fig.~\ref{Fig_scale}(a) where the same linear behavior can be seen for the ducts with very small heights $H$ and flows with very low bulk velocities and thus small Reynolds numbers. Based on the slope, the parameter $\theta_{\mathrm{SHE}} \lambda^{2} \xi_{\mathrm{lam}}^\mathrm{Rec}$ is estimated to be $1.046\times 10^{-24} \mathrm{J} \mathrm{s} \mathrm{m}^{-1}$ which is almost half of that of the circular laminar pipe flow, $\theta_{\mathrm{SHE}} \lambda^{2} \xi_{\mathrm{lam}}^\mathrm{Circ}=2.11\times 10^{-24} \mathrm{J} \mathrm{s} \mathrm{m}^{-1}$ \cite{kazerooni2020electron}. This means that $\xi_{\mathrm{lam}}^\mathrm{Circ}/\xi_{\mathrm{lam}}^\mathrm{Rec}\approx 2$ when considering  $\theta_{\mathrm{SHE}}$ and $\lambda$ to be independent of the flow and thus a cross section dependence.

\begin{figure}[h!]
	\centering
	\includegraphics[width=0.45\textwidth,trim={0cm 0cm 0cm 0cm},clip]{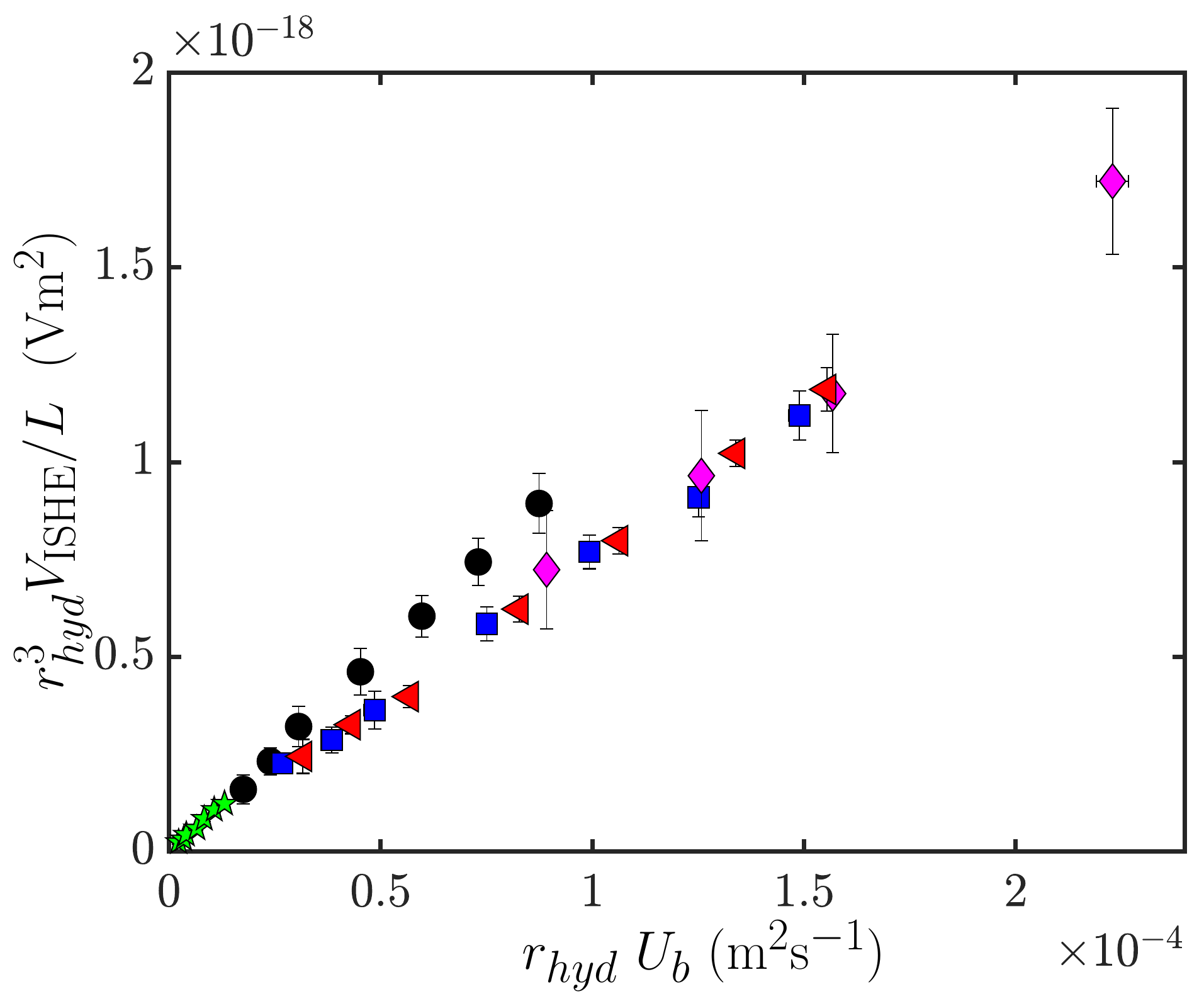}
	\caption{The generated electrical voltage $V_{\mathrm{ISHE}}$ is plotted versus $r_{hyd}U_b$ where $r_{hyd}$ is hydraulic radius. The same data are shown as in Fig.~\ref{Fig_scale}.}
	\label{Fig_scale_Hyd}
\end{figure}

In our previous study \cite{kazerooni2020electron}, we presented already one set of measured data points for the generated voltages in a laminar rectangular duct flow ($H \times W= 0.149\times 0.334~mm^2$). There, we could show that the generated voltages for this one duct case follow the same scaling law as for laminar circular pipe flows when rescaled with the hydraulic radius of the duct. However, we argued that despite this found agreement, the hydraulic diameter cannot be a universal geometrical scaling parameter for capillaries with different cross sectional shapes. Figure~\ref{Fig_scale_Hyd} brings us back to this point and replots all duct data presented in Fig.~\ref{Fig_scale}(a) now based on the corresponding hydraulic radii by using the scaling law for the laminar circular pipe flows from ref. \cite{kazerooni2020electron}. The scattering of the data points for different series clearly confirms our argument that the hydraulic radius $r_{hyd}$ as a scaling parameter for the generated voltage due to the spin hydrodynamic generation is not appropriate to capture for the geometry dependence of the electrical voltage. 

As shown in refs. \cite{kazerooni2020electron,takahashi2020giant}, it is also possible to relate the generated voltage to the pressure drop $\Delta p$ of a laminar circular pipe flow (see eq. (11) of Ref.~\cite{kazerooni2020electron}). The same analysis can be performed for rectangular duct flows and is discussed now. This results in the following correlation between the generated voltage and the pressure drop $\Delta p$ in such geometries:
\begin{equation}
\label{scaling_rect_prs}
\begin{gathered}
V_{\mathrm{ISHE}}=\frac{2|e|}{\hbar} \, \frac{\theta_{\mathrm{SHE}} \lambda^{2} \xi_{\mathrm{lam}}^\mathrm{Rec}}{ \mu \sigma_{0}} \, \Delta p\,.
\end{gathered}
\end{equation}
\begin{figure}[h]
	\centering
	\includegraphics[width=0.45\textwidth,trim={0cm 0cm 0cm 0cm},clip]{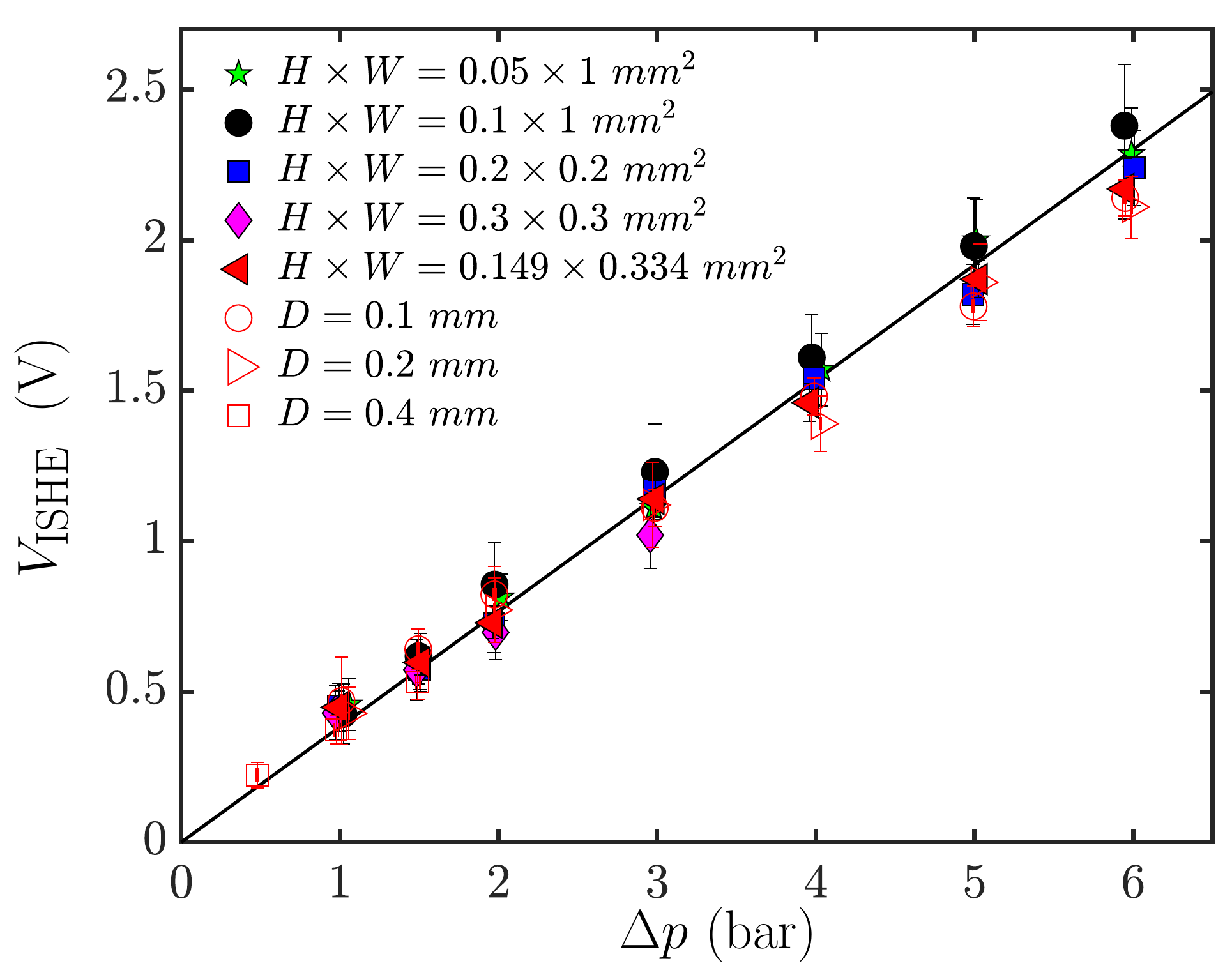}
	\caption{The generated voltage $V_{\mathrm{ISHE}}$ with respect to the impose pressure $\Delta p$ in correspondance with eq.~\ref{scaling_rect_prs}) and the corresponding counterpart for the pipe flow case that was analysed in ref.~\cite{kazerooni2020electron}.}
	\label{Fig_pressure}
\end{figure}
where $\mu$ is the dynamic viscosity of the working liquid. The results show that the electrical output of the system at the same imposed pressure is independent of the aspect ratio and extension of the duct cross section in the range considered here. Figure \ref{Fig_pressure} presents the generated voltage with respect to the imposed pressure differences ranging from 1 to 6 bar. As expected, the same voltage is generated at the same imposed pressure for different rectangular and square ducts. Interestingly, eq.~(\ref{scaling_rect_prs}) is the same as eq. (11) of Ref.~\cite{kazerooni2020electron} for circular pipe flows considering the fact that $\xi_{\mathrm{lam}}^\mathrm{Circ}\approx2\xi_{\mathrm{lam}}^\mathrm{Rec}$. This is also indicated in Fig. ~\ref{Fig_pressure} where the generated voltages in circular pipes with diameters of $D=0.1, 0.2$ and 0.4 mm are the same as in the rectangular ducts at the same pressure drop suggesting a universal behavior in the accessible range.
\begin{figure}[h]
	\centering
	\includegraphics[width=0.9\textwidth,trim={0cm 0cm 0cm 0cm},clip]{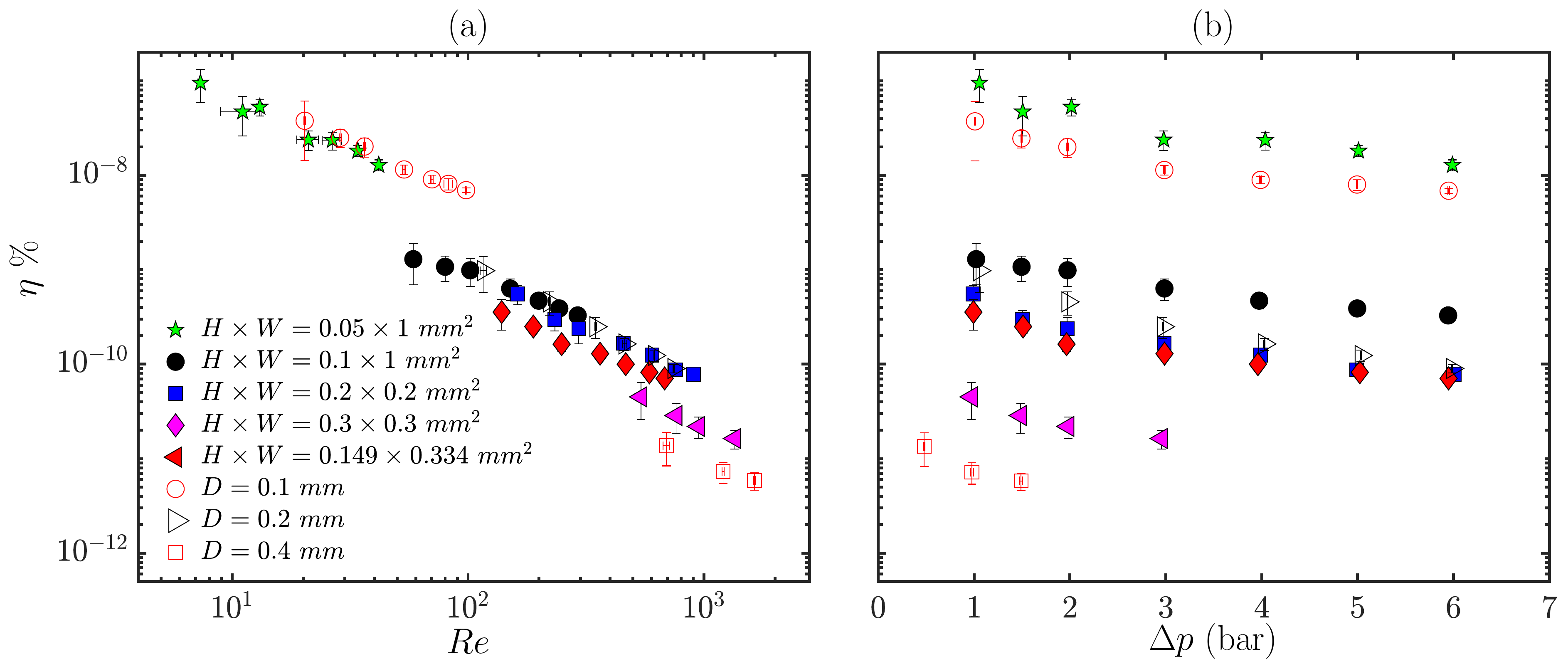}
	\caption{The system efficiency $\eta$ versus (a) the flow Reynolds number $Re$, (b) the imposed pressure $\Delta p$ for circular pipes, square and rectangular ducts.}
	\label{Eff}
\end{figure}

\subsection{Efficiency of spinhydrodynamic generation}
Finally, following \citet{takahashi2020giant}, we calculate the efficiency of the system $\eta=W_{out}/W_{in}$ defines as the ratio of the electrical power output of the system $W_{out}=A\sigma_{0}V_{\mathrm{ISHE}}/L$ to the input kinetic energy per unit time $W_{in}=A\rho U_{b}^3/2$ where $A$ is the duct cross sectional area. Here, $L$ is the length of the capillary. In Fig. \ref{Eff}(a), the calculated efficiencies $\eta$ are plotted versus the Reynolds number $Re$ in a log-log diagram for the sake of clarity. Obviously, the system efficiency is inversely proportional to the capillary tubes characteristic size and the flow Reynolds number. Using eq. (\ref{scaling_rect}) and $\eta=W_{out}/W_{in}$, it can be shown that $\eta_{\mathrm{Rec}}\propto L C_{asp}^2b^{-3}Re^{-1}$ for laminar duct flows as follows:

\begin{equation}
\label{effi_Scale_Rec}
\begin{gathered}
\eta_{\mathrm{Rec}}=\frac{144L}{\sigma_{0} \mu} C_{\mathrm{asp}}^{2}\left(\frac{|e|}{\hbar}\left(\theta_{\mathrm{SHE}} \lambda^{2} \xi_{\mathrm{lam}}^\mathrm{Rec}\right)\right)^{2} \frac{1}{b^{3}} \frac{1}{Re}
\end{gathered}
\end{equation}
In the same manner, the efficiency of a laminar pipe flow would be:
\begin{equation}
\label{effi_Scale_Circ}
\begin{gathered}
\eta_{\mathrm{Circ}}=\frac{256L}{\sigma_{0} \mu} \left(\frac{|e|}{\hbar}\left(\theta_{\mathrm{SHE}} \lambda^{2} \xi_{\mathrm{lam}}^\mathrm{Circ}\right)\right)^{2} \frac{1}{r_{0}^{3}} \frac{1}{Re}
\end{gathered}
\end{equation}
where $r_0$ is the pipe radius while the Reynolds number here is based on the pipe diameter $D$. Figure. \ref{Eff}(a) contains also the efficiency of three circular pipes with diameter of $D=0.1, 0.2$ and 0.4 mm. For a high aspect ratio rectangular duct where the geometrical coefficient $C_{\mathrm{asp}} \xrightarrow~1$, the efficiency is less than a pipe with the same diameter as the duct height considering again $\xi_{\mathrm{lam}}^\mathrm{Circ}\approx2\xi_{\mathrm{lam}}^\mathrm{Rec}$ . This can be seen from Fig. \ref{Eff}(a) where the efficiency for a pipe with $D=0.1~\mathrm{mm}$ is higher than the rectangular duct with $H \times W= 0.1\times 1~\mathrm{mm}^2$ in the same Reynolds number range. However, as the hydraulic diameters get close to each other, the efficiencies become closer at the same Reynolds number. This is clear in Fig. \ref{Eff}(a) for a pipe with $D=0.1~\mathrm{mm}$  and the rectangular duct with $H \times W= 0.05\times 1~\mathrm{mm}^2$ and a hydraulic diameter of  $D_{hyd}=2 r_{hyd}=2WH/(W+H)=0.095~\mathrm{mm}$. Similar behavior can be observed for a pipe with $D=0.2~\mathrm{mm}$ and the square and rectangular ducts with $H \times W= 0.2\times 0.2~\mathrm{mm}^2$ ($D_{hyd}=0.2~\mathrm{mm}$) and $H \times W= 0.149\times 0.334~\mathrm{mm}^2$ ($D_{hyd}=0.2 \mathrm{mm})$, respectively. The highest geometrical coefficient in eq.~(\ref{effi_Scale_Rec}) is $C_{\mathrm{asp}}=2.70$ and belongs to square ducts regardless of their height $H$. Hence, based on eq.~(\ref{effi_Scale_Rec}), (\ref{effi_Scale_Circ}) and $\xi_{\mathrm{lam}}^\mathrm{Circ}\approx2\xi_{\mathrm{lam}}^\mathrm{Rec}$, the efficiency of a square duct is the same as a circular pipe with a diameter $D$ equal to the square duct height $H$. This can also be seen in Fig. \ref{Eff}(a) for the square duct $W \times H= 0.2\times 0.2~\mathrm{mm}^2$ and circular pipe with $D=0.2~\mathrm{mm}$. Therefore, from the practical point of view and due to ease of manufacturing and the possibility to scale up the process and the system output, small microchannels with square cross-sections are suggested to use as spin hydrodynamic generators. 

It is also interesting to see the system efficiency based on the required imposed pressure for generating the flow as indicated in Fig. \ref{Eff}(b). These results show that the highest efficiency is achieved for the duct with smallest height $W \times H= 0.05\times 1~\mathrm{mm}^2$ at the lowest imposed pressure under investigation. No need to say that the efficiency in general is still too small for any practical use at the current stage.  

\section{Conclusions and outlook}
In the present study, we first derived a scaling law for the spin hydrodynamic generated voltage in a laminar duct flow by solving the spin diffusion equation analytically. It was shown that the generated voltage in a laminar duct, similar to that in a laminar circular pipe flow, is linearly proportional to the flow bulk velocity $U_b$ with a different slope. Next, the proposed scaling law was experimentally tested using different capillary tubes with square and rectangular cross sections. The measured electrical voltages were found to be in very good agreement with the predicted linear scaling law. A universal scaling for pipes and ducts in the accessible range can be obtained when we convert the dependence on the bulk velocity to pressure drop.  Moreover, we estimated the system efficiency based on the kinetic energy of the flow and the electric output of the system. Similar to a laminar pipe flow \cite{takahashi2020giant}, the efficiency was shown to be inversely proportional to the duct characteristic length, i.e. duct height, and the flow Reynolds number. We also showed that the SHDG efficiency of a square duct is equal to a circular pipe when the duct height is the same as the pipe diameter. 

As already said in the text, the voltages and efficiencies are still very small. The next question would be if a further enhancement of the spin voltage into the microvolt range is possible by further geometric modifications, such as curved channels that induce Dean vortices and nano-structured surfaces. Another option is to construct devices in which the electricity generation by SHDG can run in parallel. These studies are currently under way and will be reported elsewhere.    

\acknowledgements
This work was financially supported by the Volkswagen Foundation under the program "Experiment!". We thank Alexander Thieme for his support of the experiments. Georgy Zinchenko was supported by the German Academic Exchange Service (DAAD) with a fellowship for a double degree master program between the Moscow Power Engineering Institute (Russia) and Technische Universit\"at Ilmenau (Germany). 

\bibliography{Paper}
\end{document}